\def\aa{{A\&A}}

\def\apj{{ApJ}}

\documentclass[cup5b]{caps}
\usepackage{graphicx}
\usepackage{amssymb}
\usepackage{ociwsymp4e}

\HeadText{S. W. Campbell}

\def\plottwo#1#2{\centering \leavevmode
\includegraphics[width=.6\columnwidth]{#1} \hfil
\includegraphics[width=.6\columnwidth]{#2}}

\begin{document}

\pagenumbering{arabic}

\author[]{S. W. CAMPBELL\\Centre for Stellar and Planetary Astrophysics, Monash University, Australia}

\chapter{Nucleosynthesis in Early Stars}

\begin{abstract}

We present preliminary results of stellar structure and nucleosynthesis calculations for some early stars. The study 
(still in progress) seeks to explore the expected chemical signatures of second generation low- and intermediate-mass 
stars that may have formed out of a combination of Big Bang and Pop III (Z=0) supernovae material. Although the 
study is incomplete at this stage, we find some important features in our models. The initial 
chemical composition of these early stars is found to be significantly different to that given by just scaling the solar 
composition. The most notable difference is the lack of nitrogen. This should not affect the structural evolution 
significantly as nitrogen will be quickly produced through the CNO cycle due to the presence of carbon (and oxygen). It should however
effect the nucleosynthetic yields. We also find that our very low metallicity $5M_\odot$ model, with [Fe/H] $= -4.01$, 
does {\em not} reach the RGB --- it goes directly to the helium burning phase. It does not experience the first dredge-up either. 
This is not a new finding but it will have an effect on the surface chemical evolution of the stars and should alter the nucleosynthetic yields that 
we are currently calculating. Our higher metallicity stars, with a globular cluster composition at [Fe/H] $= -1.4$, do experience all the
standard phases of evolution but also have significantly higher surface temperatures and luminosities compared to solar
metallicity stars. Their internal temperatures are also higher which should again effect the final nucleosynthetic yields.

\end{abstract}

\section{Introduction}

The first stars formed out of the material synthesised by the Big Bang. Big Bang Nucleosynthesis (BBN) theory 
makes quantitative predictions about the nuclei that were formed just after the Big Bang. The 
material was primarily hydrogen and helium, with some heavier trace elements. As the heavier trace elements produced in standard BBN do not 
have a significant effect on stellar structure, many researchers have computed stellar models with a Z=0 (pure H
 and He) starting composition to represent the first generation of stars (eg. Weiss, et al. 1999, Chieffi, et al. 2001, Siess, 
Livio, \& Lattanzio 2001).

The mass distribution of these stars, otherwise known as the First Mass Function (FMF), is central to tracing the 
chemical history of the Universe, as the stellar nucleosynthetic yields that pollute the Universe are 
primarily dependent on the mass of the stars. The nature of the FMF has always been a contentious issue and remains
unresolved. However, recent 3-dimensional hydrodynamical star formation simulations (at Z=0, see Abel, Bryan, \&  Norman 2000) 
support the theories that predict a FMF biased towards high mass stars. Abel et al. (2000) predict star formation 
masses of the order $10^2M_\odot$. This is due to the different formation environment and different protostellar cloud
composition (as compared to later star-forming epochs). As massive stars have very short lifetimes they would have 
already ended their evolution and would therefore be 
unobservable today (at least locally). However, it has also been suggested that the FMF may have been {\em bimodal}, 
with one peak at roughly $1M_\odot$, and the other at higher masses (of the order $10^2M_\odot$, see Nakemura \& Umemura, 2001). 
If there were a population of low-mass stars, some may still exist --- as their lifetimes are comparable to the age of the Universe. 
Ultra-Metal-Poor Halo stars (UMPHs) are possible candidates for these first stars but may also be members of a second 
generation of stars.

Due to the uncertainty of the nature of the formation of the first population of stars (Population III), and the resulting
uncertainty in the FMF, we have decided to take two different approaches to determining the initial composition of 
our second generation models. These scenarios are described in the next two subsections. For both scenarios, due 
to the abnormal composition and low metallicity of the stars, it has been necessary to obtain custom-made opacity 
tables (from OPAL) and increase the temperature range in which the Mt. Stromlo stellar structure and evolution code
$EVOLN$ operates. Mass loss is included in the simulations - Reimers' (1975) relation on the RGB and Vassiliadis 
and Wood's (1993) on the AGB. The code includes structurally important nuclear reactions only. Also see  Wood \& Zarro 
(1981) and Frost \& Lattanzio (1996) for further details on this code. 
After the structural evolution is completed, we use a separate code to calculate the detailed nucleosynthesis. The Monash 
University code $DPPNS45$  reads in the physical structure output from $EVOLN$ and computes the nucleosynthesis for 
the lifetime of the star using a reaction network of 74 isotopes and 506 reactions.

\subsection {Scenario One: Big Bang Material Mixed with a Pop III Supernova}

Here we create a model star that formed out of the debris of a $20M_\odot$, Z=0 supernova
(using yields from Limongi et al, 2002, private communication). We mixed the yield with Big Bang
material down to a metallicity of [Fe/H] $= -4.01$ ($Z=10^{-6}$). This metallicity is indicative of that
found in the UMPHs. It took $10^6$ solar masses of Big Bang material to achieve this metallicity.
The rationale behind this scenario is that the first generation of stars were 
probably massive and would have exploded as supernovae and quickly polluted the ISM.
The shocks from the first SNe probably triggered the formation of the next generation of stars.
Star formation material for the second generation of stars may have only 
been polluted by a few (or even one) of these events.

\subsection {Scenario Two: Second Generation Globular Cluster Stars}

This scenario involves feedback from a globular cluster chemical evolution (GCCE) model currently being computed 
by Fenner et al. (Swinburne University). They are attempting to explain some of the isotopic anomalies
in NGC 6752 (eg. the Mg-Al anticorrelation, Gratton et al. 2001). Their GCCE model assumes a Prompt Initial 
Enrichment of the protocluster by a generation of zero metallicity Population III stars. The Z=0 
stellar yields used for the GCCE are also from Limongi (private communication). A top-heavy mass distribution is used
for the protocluster. The resulting second generation globular cluster stars form with a metallicity of [Fe/H] $= -1.4$, 
which is the observed metallicity of NGC 6752. However, the composition of the formation gas is 
non-scaled-solar. Thus they require a feedback of stellar yields computed with the abnormal composition so that the 
model is self-consistent.


\section{Results}

\subsection{Scenario One: SN + BBN}

\begin{figure}
    \centering
    \includegraphics[width=10cm,angle=0]{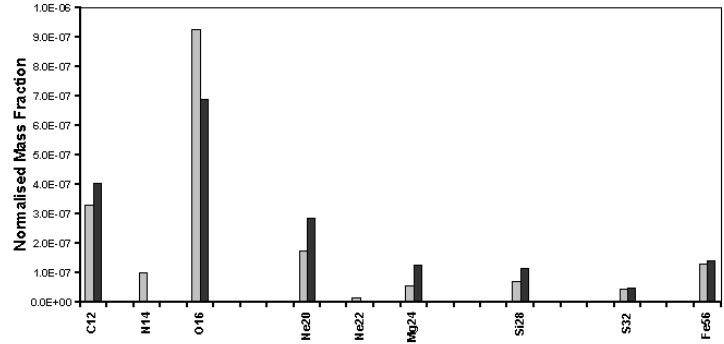}
    \caption{Part of the initial composition of the SN ejecta + BBN mix stellar model (dark grey), as compared
             to a scaled solar composition with the same [Fe/H] $= -4.01$ (light grey). The main
             difference is the lack of Nitrogen in the SN + BBN model.}
    \label{snbbcomp}
  \end{figure}

Selected isotopes of the resulting composition for the SN + BBN composition are shown in Figure~\ref{snbbcomp}. 

The most notable features in the initial composition are the lack on nitrogen and the overabundance of some 
of the $\alpha$~-elements (as compared to scaled Solar). The almost total lack of nitrogen will probably have little 
effect on the structural evolution of the star as nitrogen will be produced in the core very quickly during the main sequence.
However, it should have an effect on the final yields. 

Figure~\ref{singles} shows the first stellar model we have computed using this composition. A track for a 
solar metallicity star of the same mass is also shown for comparison. Some important features are:

\begin{itemize}
\item{There is no Red Giant Branch --- it doesn't get near its Hyashi limit.}
\item{There is no First Dredge-up --- this will affect the surface composition. }
\item{The surface temperature and luminosity are much higher than the solar 
metallicity star.}
\end{itemize}

The calculation has not yet been taken through to the end of the AGB so we have no 
nucleosynthetic yields so far.

\subsection{Scenario Two: GCCE}

\begin{figure}
    \centering
    \plottwo{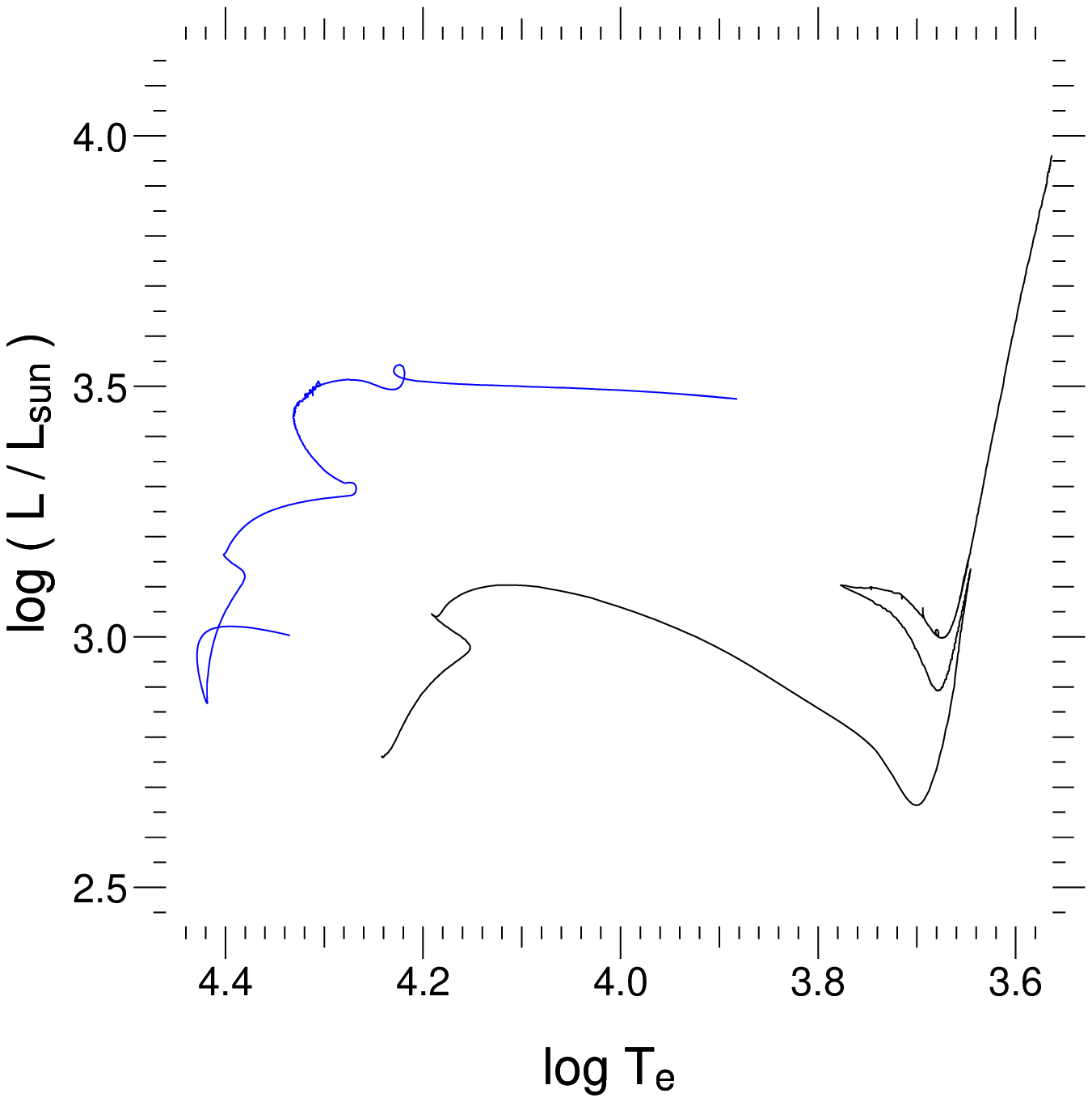}{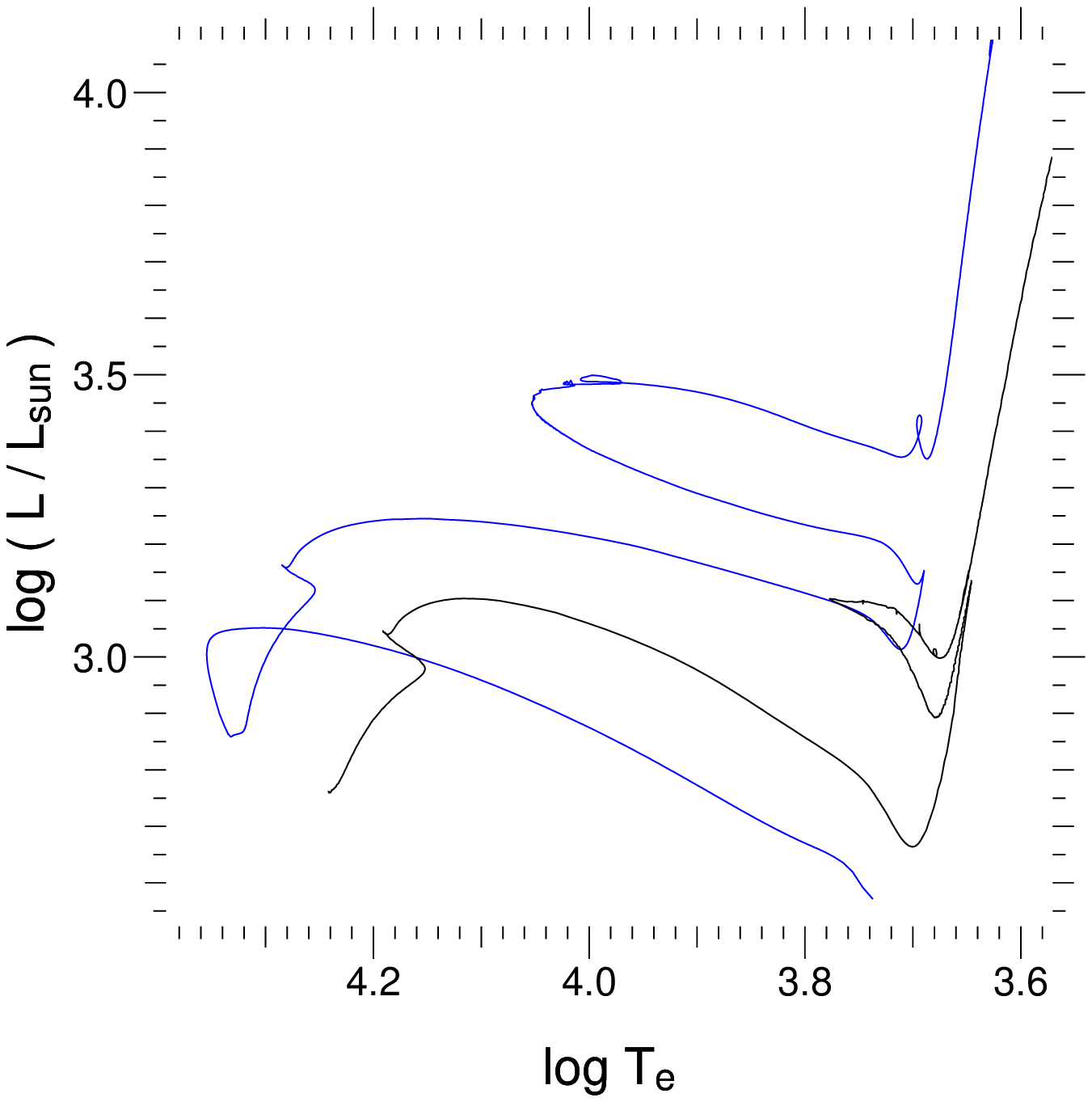}
    \caption{{\em Top:}  Comparison in the HR diagram between a $5M_\odot$ star of solar metallicity (lower track) 
      and our very low metallicity ([Fe/H] $= -4.01$, $Z=10^{-6}$) BB + SN star, also with a mass of $5M_\odot$ (upper track).
      The important differences are highlighted in the main text.  {\em Bottom:} Comparison in the HR diagram between a 
      $5M_\odot$ star at solar metallicity (lower track) and the $5M_\odot$ GC star (upper track). Even at this metallicity 
      ([Fe/H] $= -1.4$), the GC star has a significantly higher luminosity and surface temperature.}
    \label{singles}
  \end{figure}

\begin{figure}
    \centering
    \plottwo{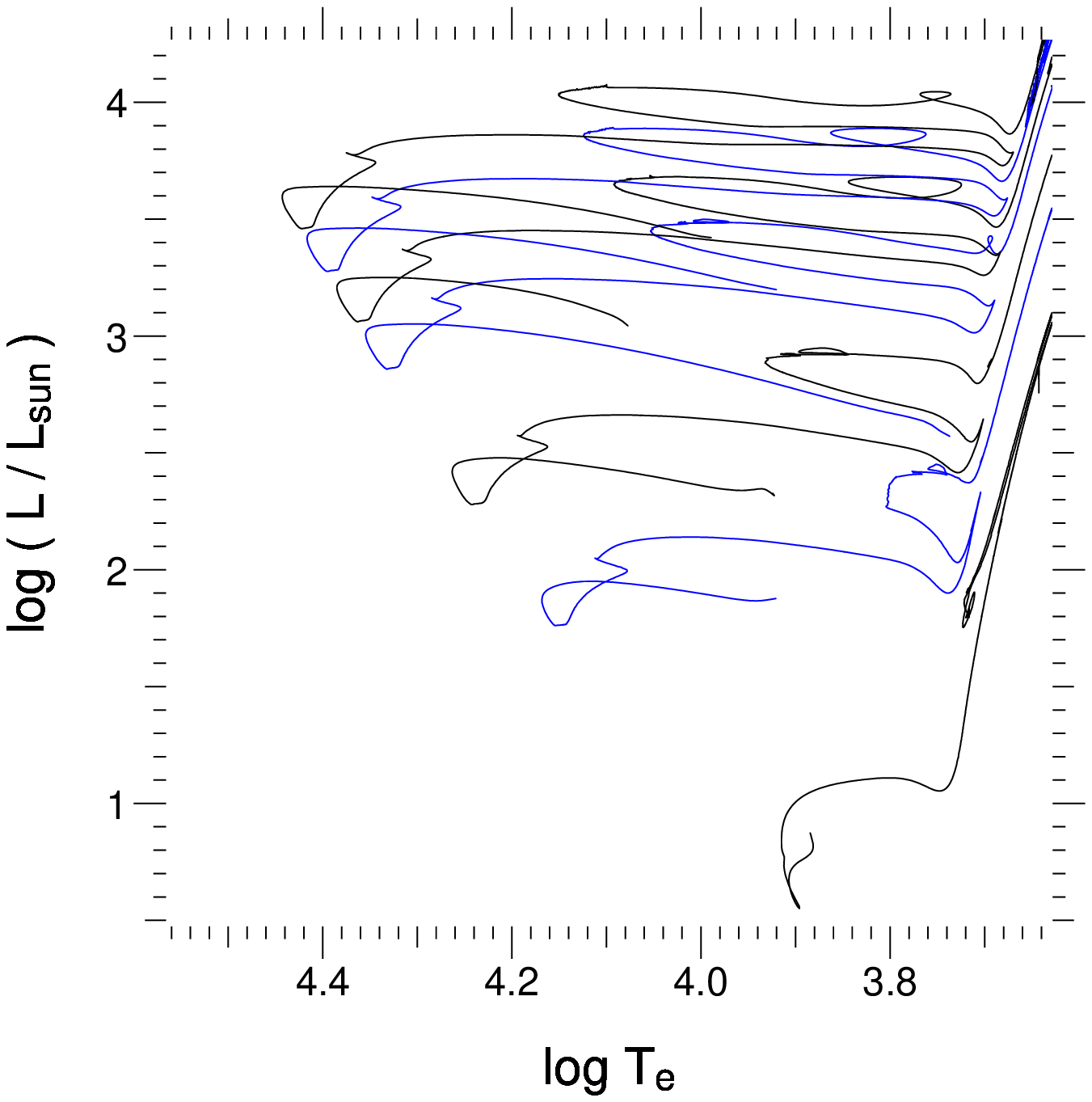}{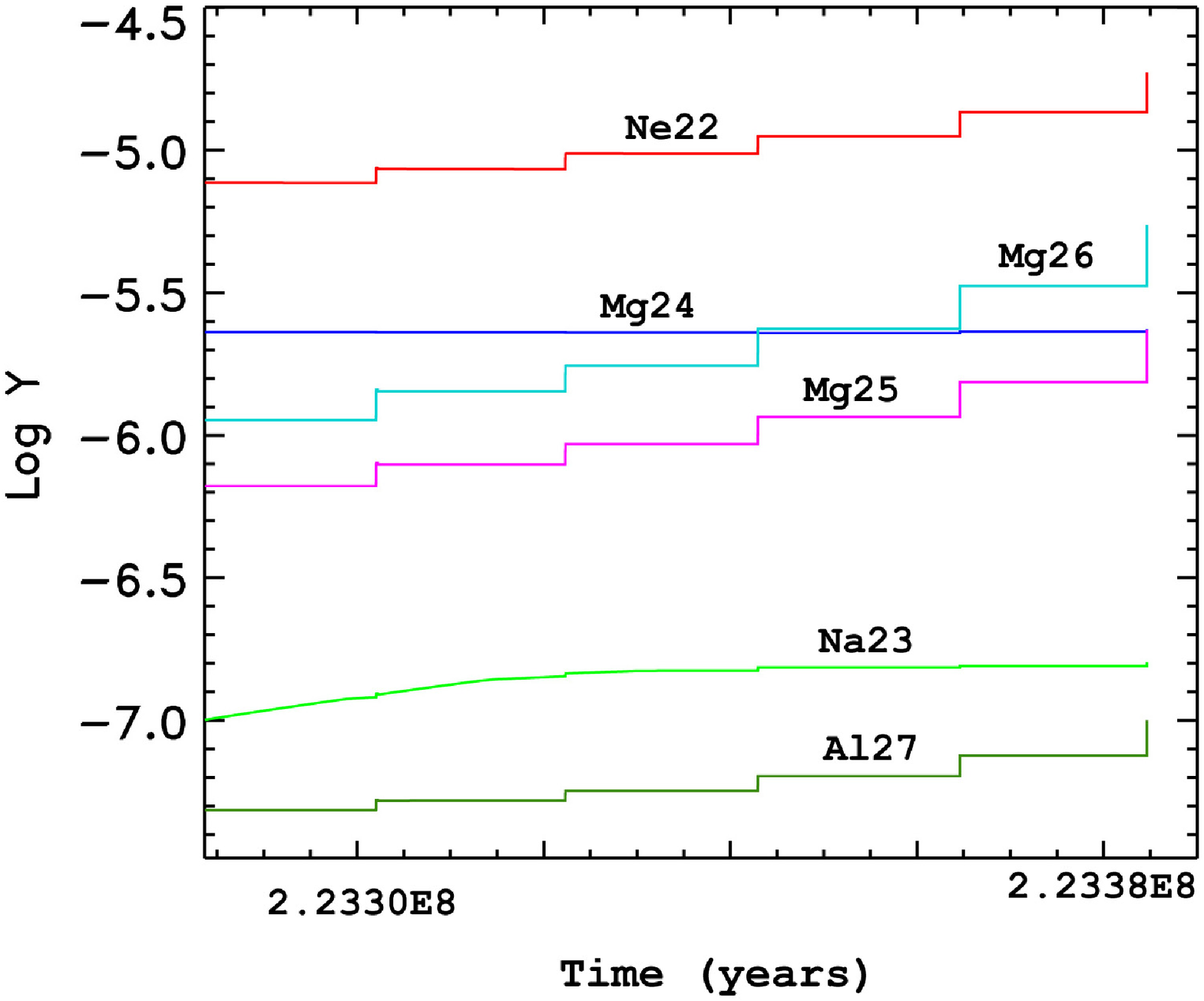}
    \caption{{\em Top:} HR diagram. All the non-scaled-solar GC models computed so far. From bottom to top the initial masses are: 1.25, 2.5, 3.5, 5.2, 6.0, 7.0, $8.0 M_\odot$ {\em Bottom:} End of the evolution of the surface composition of the $3.5M_\odot$ GC star, for some selected isotopes. The 
      periodic enrichment occurs through the repeated dredge-up of processed material from the intershell region during 
      the thermally-pulsing AGB stage.}
    \label{gchrdall}
  \end{figure}

We are currently computing a grid of low- and intermediate-mass (~$1-7M_\odot$) models 
for the Fenner et al. NGC 6752 model. The initial composition has some of the same salient features as that of 
scenario 1, although  the most notable feature is again the lack of nitrogen.
Most of the stellar structure calculations have been completed. Figure~\ref{singles} is shown as an example
to highlight the differences between the evolution of a solar-metallicity star and a low-Z globular cluster (GC) model. Unlike the very low
metallicity BBN + SN star of scenario 1, the GC star does reach the RGB and experiences the first 
dredge-up. Figure~\ref{gchrdall} shows the tracks of all the models done to date, in the HR diagram.

Nucleosynthesis calculations have also been completed for some of the low-mass models. As an example,
Figure~\ref{gchrdall} shows the periodic enrichment of the surface during successive Thermal Pulses on the AGB.
The final two pulses have been computed synthetically as the stellar code had convergence difficulties at the 
end of the AGB (see Karakas and Lattanzio, 2003). This method gives more complete yield approximations for feedback into the GC chemical 
evolution model. This is the only synthetic modelling used and is only necessary for the last few pulses in some of the models.


\section{Conclusions}

The current study is a work in progress. However, some important features of the results so far can be discussed.

The lack of nitrogen arises because it is not synthesised in (standard) BBN nor in any significant amount in the 
Z=0 massive supernovae used for our starting compositions. The quantitative consequences of this and other 
effects are yet to be determined but it will probably have the most effect on the final yields. It must be noted however that 
this lack of nitrogen is model-dependent --- some Z=0 models do in fact produce nitrogen. This needs further consideration 
--- nitrogen may need to be a free parameter in the study. The discrepancy in Z=0 nitrogen yields arises from the
uncertainty in reaction rates in those models. 

In general, our low metallicity stars evolve differently to solar metallicity stars. They have hotter, more compact cores --- 
due to different nuclear burning. Their surface temperatures and luminosities are also higher. In addition, our $5M_\odot$ 
very low metallicity star did not experience the first dredge-up, keeping the surface pure during the horizontal branch stage. 
These features are not new findings but they do support work done previously by others. They are also likely to alter 
the final nucleosynthetic yields of the stars, particularly if the extreme temperatures
persist into the thermally-pulsing AGB phase. We have recently found temperatures at the base of the convective envelope exceeding
$10^8K$ ($7M_\odot$ model) - this will produce hot bottom burning and thus the abundances of elements such as lithium
on the surface will be altered.

The fact that it took $10^6M_\odot$ of Big Bang material to get a starting metallicity of the SN + BB model comparable to 
that of the lowest metallicity Halo stars is also interesting. If the UMPHs are very old second generation stars, this figure
may give an indication of the size of the original protostellar cloud from which these stars formed.

Further investigation of Big Bang Nucleosynthesis is needed, as the evolution of the first stars is very sensitive to trace 
elements, in particular carbon. Was carbon produced in the Big Bang? This would have a significant effect on Population 
III stellar evolution and hence the starting composition for the next generation.

The final stage of the study will be to compare our results with observations. Can we match the observed chemical 
signatures in UMPHs or GC stars with our theoretical yields?


\end{document}